\documentclass[a4paper,12pt]{article}
\usepackage{amssymb,latexsym,amsfonts,amstext,amsmath}

\title{Von Neumann's book,\thanks{Mathematische Grundlagen der
Quantenmechanik, 1932}\,\, the Compton-Simon experiment and the collapse
hypothesis}

\author{R. N. Sen\\[1mm]
Department of Mathematics\\ 
Ben-Gurion University of the Negev\\
Beer Sheva 84105, Israel}

\begin{document}

\maketitle

\thispagestyle{empty}

\begin{abstract}

Few things in physics have caused so much hand-wringing as von
Neumann's collapse hypothesis. Unable to derive it mathematically, von
Neumann attributed it to interaction with the observer's brain! Few
physicists agreed, but tweaks of von Neumann's measurement theory did
not lead to collapse, and Shimony and Brown proved theorems
establishing `the insolubility of the quantum measurement problem'.
Many different `interpretations' of quantum mechanics were put
forward, none gained a consensus, and some scholars suggested that the
foundations of quantum mechanics were flawed to begin with. Yet, for
almost ninety years, no-one looked into how von Neumann had arrived at
his collapse hypothesis!

Von Neumann based his argument on the experiment of Compton and Simon.
But, by comparing readings from von Neumann's book and the
Compton-Simon paper, we find that the experiment provides \emph{no
evidence} for the collapse hypothesis; von Neumann had misread it
completely! We suggest that von Neumann had relied on his phenomenal
memory rather than the printed Compton-Simon paper, and his memory had
failed him for once. Our finding has considerable implications for physics,
which -- briefly sketched here -- will be discussed elsewhere in
detail. An Appendix raises some questions for historians of physics.

\end{abstract}

\pagebreak

\section{Introduction}\label{INTRO}

In his book \emph{Mathematische Grundlagen der Quantenmechanik}
(published 1932) \cite{VN1932}, von Neumann made an assertion that has
spooked the world of physics ever since: the state vector evolves with
time in \emph{two different ways}: when left alone, it changes
smoothly and reversibly, according to Schr\"odinger's equation; but,
upon measurement of an observable, it \emph{collapses}, abruptly and
irreversibly, into an eigenvector of the observable.\footnote{Worse,
the collapse happens only when the observation reaches the observer's
brain! Von Neumann came to this conclusion because the state vector of
his object-plus-apparatus `combined system' failed to collapse in his
measurement scheme.} Von Neumann discerned `three degrees of causality
or non-causality' in a quantum-mechanical measurement: (i)~the initial
and final states are linear combinations of eigenvectors of the
measured observable; (ii)~the initial state is a linear combination of
eigenvectors, but the final state is a single eigenvector;\footnote{We
are assuming that the eigenvalues are non-degenerate, which is a
harmless simplification.} (iii)~both initial and final states are
eigenvectors -- the measurement process is strictly causal.  He then
went on to claim that `\emph{The Compton-Simon experiment shows that
only the second case is possible in a statistical theory}'.

This last assertion of von Neumann has remained unchallenged for
ninety years.  However, a side-by-side reading of von Neumann's book
and Compton and Simon's paper \cite{CS1925} reveals that \emph{it is
simply not true}. The Compton-Simon experiment showed that energy and
momentum were conserved in the scattering of an electron by an $x$-ray
quantum, but gave no evidence that any state vector had collapsed.
Indeed it could not, for \emph{experimental} detection of collapse (or
its absence) requires not one but \emph{two} (successive) measurements.
Somehow, von Neumann had got his physics wrong!

In the next Section, we shall justify the above assertion by comparing
readings from von Neumann's book and the Compton-Simon paper. In
Sec.~\ref{MEMORY-FAIL}, we shall try to understand how von Neumann
fell into this error. His collapse hypothesis -- more precisely, its
unacceptability -- has given rise to several `interpretations' of
quantum mechanics; our questioning of it will produce yet another one,
with a greater emphasis on physics. It will, hopefully, be the subject
of a separate memoir in a physics journal, but we shall provide a
summary of it in Sec.~\ref{PHYSICS}; the present work prepares the
background for this interpretation. 

Some unanswered questions and two private communications which may be
of interest to the historian of physics will be presented in the
Appendix.

%%%%%%%%%%%%%%%%%%%%%%%%%%%%%%%%%%%%%%%%%%%%%%%%%%%%%%%%%%%%%%%%%%%

\section{Readings from von Neumann's book and and Compton and Simon's
paper}\label{STAGGERING}

We begin with what von Neumann wrote on pp~212--213 of his
book:\footnote{In the \emph{Translator's Preface} (to the English
translation), dated December 1949, the translator (R. T. Beyer) says:
`The translated manuscript has been carefully revised by the author so
that the ideas expressed in this volume are his rather than those of
the translator, and any deviations from the original text are also due
to the author.'  {\bf Our page references are to the original (1955)
edition of the Beyer translation}.}

\begin{quote}

First, let us refer to an important experiment which Compton and Simon
carried out prior to the formulation of quantum
mechanics.\footnote{Footnote in the original: Compton and Simon
\cite{CS1925}. Cf.\ the comprehensive treatment of W.  Bothe in
\cite{BOTHE1926}, Chapter 3 in particular, \& 73.\label{123}} In this
experiment, light was scattered by electrons, and the scattering
process was controlled in such a way that the scattered light and the
scattered electrons were subsequently intercepted, and \emph{their
energy and momenta measured} [our emphasis]. That is, there ensued
collisions between light quanta and electrons, and the observer, since
he measured the paths after collision, could prove whether or not the
laws of elastic collision were satisfied\ldots The experiment gave
complete confirmation to the mechanical laws of collision
[conservation of energy and momentum].

\end{quote}

\noindent We have omitted a part of the paragraph before the last
sentence. This omission does not affect the meaning of the quotation.
The important part is von Neumann's assertion that \emph{the energy
and momenta of the recoil electron and the scattered photon were
measured} in the Compton-Simon experiment! This assertion is factually
incorrect; energy and momenta of the recoil electron and the scattered
photon were \emph{not} measured in the experiment, as we shall see
from a reading of the paper.

The experiment of Compton and Simon was designed to test the
hypothesis, advanced in 1924 by Bohr, Kramers and Slater (hereafter
BKS) to avoid the need for radiation quanta \cite{BKS1924}, that
energy and momenta were conserved only statistically, and not in
individual collisions between electrons and radiation. Measurement of
the energy-momenta of the recoil electron and scattered photon would
undoubtedly have been sufficient, but \emph{was not necessary} for
this test; it would have been enough to determine the angles $\theta$
and $\phi$ of eq.\ (1) below.

This is what Compton and Simon wrote in their article (\cite{CS1925},
p~290):

\begin{quote}

\ldots The change of wave-length of $x$-rays when scattered and the
existence of recoil electrons associated with $x$-rays, it is true,
appear to be inconsistent with the assumption that $x$-rays proceed in
spreading waves if we retain the principle of the conservation of
momentum.\footnote{Footnote in the original: Compton
\cite{COMPTON1924}.} Bohr, Kramers and Slater,\footnote{Footnote in
the original: Bohr, Kramers and Slater \cite{BKS1924}} however, have
shown that both these phenomena and the photo-electric effect may be
reconciled with the view that radiation proceeds in spherical waves if
the conservation of energy and momentum are interpreted as statistical
principles.

A study of the scattering of individual $x$-ray quanta and of the
recoil electrons associated with them makes possible, however, what
seems to be a crucial test between the two views of the nature of
scattered $x$-rays.\footnote{Footnote in the original: The possibility
of such a test was suggested by W. F. G. Swann in conversation with
Bohr and one of us in November 1923.} On the idea of radiation quanta,
each scattered quantum is deflected through some definite angle $\phi$
from its incident direction, and the electron which deflects the
quantum recoils at an angle $\theta$ given by the
relation\footnote{Footnote in the original: Debye \cite{DEBYE1923};
Compton and Hubbard \cite{CH1924}}
\begin{equation}
\tan {\textstyle{\frac12}}\phi = - \frac{1}{[(1+\alpha)\tan \theta]}
\end{equation}
where $\alpha = h/mc\lambda$. [Here $\lambda$ is the wavelength of the
incident $x$-ray.] Thus a particular scattered quantum can produce an
effect only in the direction determined at the moment it is scattered
and predictable from the direction in which the recoiling electron
proceeds. If, however, the the scattered $x$-rays consist of spherical
waves, they may produce effects in any direction whatever, and there
should consequently be no correlation between the direction in which
recoil electrons proceed and the directions in which the effects of
the scattered $x$-rays are observed.

\end{quote}

Compton and Simon summarise their main numerical data in the Abstract
of their paper (p~289):

\begin{quote}

Of the last 850 plates, 38 show both recoil tracks and
$\beta$-tracks'.  [The electrons scattered by the deflected photons
are called `$\beta$-electrons' by the authors]\ldots in 18 cases, the
direction of scattering is within $20^{\circ}$ of that to be expected
if\ldots energy and momentum are conserved during the interaction
between the radiation and the recoil electron. This number 18 is four
times the number which would have been observed if the energy of the
scattered $x$-rays proceeded in spreading waves\ldots The chance that
this agreement with theory is accidental is about 1/250. The other 20
$\beta$-ray tracks are ascribed to stray $x$-rays and to
radioactivity\ldots

\end{quote}

They conclude their paper as follows (p~299):

\begin{quote}

These results do not appear to be reconcilable with the view of the
statistical production of recoil and photo-electrons proposed by Bohr,
Kramers and Slater.  They are, on the other hand, in direct support of
the view that \emph{energy and momentum are conserved during the
interaction between radiation and individual electrons} [emphasis in
the original].

\end{quote}

Let us return to the experiment itself.  A nearly monochromatic
$x$-ray beam was scattered by the air in an expansion cloud chamber,
and the event photographed.  Most of the plates recorded only noise,
but a few showed (i)~the points of ejection of the recoil electron and
that of the $\beta$-electron, and (ii)~the direction of recoil of the
recoil electron. These sufficed to determine the angles $\theta$ and
$\phi$.  If energy and momentum were conserved in the initial
collision, $\theta$ and $\phi$ would satisfy eq.~(1); otherwise not
(the wavelength $\lambda$ of the incident $x$-ray beam was known).
There was no need to determine the energy-momenta of either the recoil
electron or of the scattered $x$-ray quantum, nor was it attempted by
the authors.

That is, Compton and Simon's paper does not support von Neumann's
remark, quoted above, that `the scattered light and the
scattered electrons were subsequently intercepted, and their energy and
momenta measured\ldots'

Von Neumann goes on to infer the following (op.\ cit., p 213):

\begin{quote}
More generally, the experiment shows that the same physical quantity
(namely, any coordinate of the place of collision or of the direction
of the central line [defined earlier as the direction of momentum
transfer by von Neumann]) is measured in two different ways (by
capture of the light quantum and of the electron), and the result is
always the same.

These two measurements do not occur entirely simultaneously. The light
quantum and the electron do not arrive at once, \emph{and by suitable
arrangement of the measuring apparatus either process may be observed
first} [our emphasis]. The time difference is usually about $10^{-9}$
to $10^{-10}$ seconds [the time it takes a photon to traverse three to
30 cms]\ldots
\end{quote}

The rest of the paragraph adds nothing new, and has been left out.
Again, \emph{the assertions in the paragraph are completely wrong, and
bear no relation to the experiment of Compton and Simon}.

In the next two paragraphs (op.\ cit., pp 213--214), von Neumann makes
a far-reaching generalisation of his misconceptions (denoting by
$\mathcal{R}$ the physical quantity being measured). 

\begin{quote}
We can formulate the principle that is involved as follows: by nature,
three degrees of causality or non-causality may be distinguished.
First, the $\mathcal{R}$ value could be entirely statistical, i.e.,
the result of a measurement would be predicted only statistically; and
if a second measurement were then taken immediately after the first
one, this would also have a dispersion, without regard to the value
found initially -- for example, its dispersion might be equal to the
original one.\footnote{Footnote in the original: A statistical theory
of elementary processes was erected by Bohr, Kramers and Slater on
these basic concepts. Cf.\ \cite{BKS1924} and references cited in
[footnote \ref{123}]. The Compton-Simon experiment can be considered
as a refutation of this view.\label{FOOTNOTE-CS}} Second, it is
conceivable that the value of $\mathcal{R}$ may have a dispersion in
the first measurement, but that \emph{immediately subsequent
measurement is constrained to give a result which agrees with that of
the first} [our emphasis].  Third, $\mathcal{R}$ could be determined
causally at the outset.

The Compton-Simon experiment shows that only the second case is
possible in a statistical theory\ldots

\end{quote}

Observe carefully the generality of the above assertion; von Neumann
is referring to the measurement of an arbitrary `physical quantity'
(self-adjoint operator) $\mathcal{R}$!

Von Neumann's misunderstanding of the Compton-Simon experiment was
complete; in (our) footnote \ref{FOOTNOTE-CS}, he asserts: `The
Compton-Simon experiment \emph{can be considered} [our emphasis] as a
refutation of this [the BKS] view'. The phrase `can be considered'
is at odds with the fact that the Compton-Simon experiment was
performed expressly to test the BKS hypothesis; refutation of the
hypothesis by the experiment was \emph{not} an afterthought.

Von Neumann's reference to \& 73 of Bothe's \emph{Handbuch} article is
equally misleading. The title of \& 73 is `The theory of Bohr,
Kramers and Slater and its experimental test'. The first part of the
article explains the BKS theory. The second part describes the
experiment of Bothe and Geiger.  The experiment is based on a
\emph{coincidence counter} designed by Geiger in which one counter
records the arrival of the recoil electron and the other that of the
scattered photon. The difference between their times of flight is
smaller than the dead times of the counters, and \emph{there is no way
in which the results can be interpreted as two separate experiments}.

The Compton-Simon experiment is relegated to the penultimate paragraph
of \& 73, which consists of the single sentence

\begin{quote}

Compton and Simon have concluded, from their researches with a Wilson
cloud chamber, that the direction of the recoil electron and that of
the scattered radiation are indeed related by eq.\ (63) [which is the
same as eq.\ (1) of Compton and Simon].

\end{quote}

%%%%%%%%%%%%%%%%%%%%%%%%%%%%%%%%%%%%%%%%%%%%%%%%%%%%%%%%%%%

\section{Did von Neumann's memory fail for once?}\label{MEMORY-FAIL}

How did von Neumann get it so wrong? Any answer would be conjectural.
One may assume that von Neumann \emph{did not consult} printed
versions of the Compton-Simon and Bothe articles when he wrote what
was quoted above. It is possible that he relied on his memory -- which
seldom failed him, but did fail him this once.  A sketch of von
Neumann's travels and his works from 1926 to 1932 (pieced together
from Macrae's biography \cite{NM1992} and the obituary by Ulam
\cite{SU1958}), and accounts of his prodigious memory, will show that
this is a possible, indeed plausible explanation.

Von Neumann arrived in G\"ottingen in early autumn, 1926. Quantum
mechanics had just been discovered there, and Hilbert's physics
assistant, Lothar Nordheim, could not explain it to him. After talking
to Nordheim, von Neumann wrote up an account which was much
appreciated by Hilbert.  A joint paper with Hilbert and Nordheim
followed in 1927. Von Neumann published six papers in 1927, of which
four were on quantum mechanics.  He was appointed Privatdozent in
Berlin in autumn 1927. In 1928, he published 10 papers, including
three with Wigner. In autumn 1929, he moved briefly to Hamburg as
Privatdozent, but left for Princeton in January 1930, where he stayed
for the rest of his life.  In 1929, he published ten papers (two of
them with Wigner).  This dropped to only one in 1930, picked up to six
in 1931 (including the one on the uniqueness of Schr\"odinger
operators), and to nine papers and his book \emph{Mathematische
Grundlagen der Quantenmechanik} in 1932.

What explains the drop in output in 1930? For one, von Neumann married
Marietta Kovesi on New Year's day, 1930, in Budapest (despite having
noticed, according to Macrae, how much she spent on clothes on an earlier
trip to Paris). The couple sailed immediately for New York from
Cherbourg, via Paris again. So, despite Bethe's musings whether ``a
brain like von Neumann's does not indicate a species superior to
man'', von Neumann was human after all, at least in some ways.

Von Neumann's memory was the stuff of legends. Two are quoted below.
The first is from Macrae's biography \cite{NM1992}, p~52:

\begin{quote}
The older of Johnny's brothers, Michael, recalled that Johnny read
through all forty-four volumes of [William Oncken's \emph{Allgemeine
Ge\-schi\-chte}]\ldots[and] what Johnny read, Johnny remembered.
Decades later friends were startled to discover that he remembered
still. He could recite whole chapters verbatim.

\end{quote}

The second, by Hermann Goldstine (von Neumann's collaborator on the
computer project in Princeton), is from the Wikipedia page
on John von Neumann \cite{WIKIPEDIA}:

\begin{quote}

One of his remarkable abilities was his power of absolute recall. As
far as I could tell, von Neumann was able on once reading a book or
article to quote it back verbatim; moreover, he could do it years
later without hesitation. He could also translate it at no diminution
in speed from its original language into English. On one occasion I
tested his ability by asking him to tell me how A Tale of Two Cities
started. Whereupon, without any pause, he immediately began to recite
the first chapter and continued until asked to stop after about ten or
fifteen minutes.

\end{quote}

However, von Neumann's memory did have some limits. We return to
Mac\-rae's biography \cite{NM1992}, p~152:

\begin{quote}

His memory and feel for words, plus unsurpassed feel for mathematical
symbols, had not extended to memory for faces. All his life he was
embarassed by not knowing people who clearly knew him. He had no sort
of photographic memory\ldots

\end{quote}

If von Neumann indeed relied on his memory -- rather than consult
printed texts, for whatever reason -- his mistakes could be explained
by a rare memory lapse. His memory probably would not have failed him
in mathematics, but physics could have been different.  Macrae reports
that Ulam once said that ``Johnny\ldots did not have a penchant
for guessing what happens in given physical situations''
(\cite{NM1992}, p.~236). The ability to rely on his memory (rather
than having to consult printed sources) for much of his work could
also explain, in part, how he could accomplish so much in so short a
time.  

%%%%%%%%%%%%%%%%%%%%%%%%%%%%%%%%%%%%%%%%%%%%%%%%%%%%%%%%%%%%%%%%%%

\section{Implications for physics}\label{PHYSICS}

Although von Neumann's collapse hypothesis was too general, it did
point in the right direction. One may recall that his example
concerned the measurement of conserved quantities. Now, the
measurement of an \emph{additively conserved quantity} on a single,
isolated object will necessarily cause the latter's state vector to
collapse to an eigenvector belonging to the measured eigenvalue; if it
does not, a subsequent measurement(on the same object) may yield a
different eigenvalue, \emph{which will violate the conservation law}.
In this case collapse is not a hypothesis; it is a consequence of the
conservation laws. The situation is reminiscent of the derivation of
the Boltzmann equation from Newton's laws. The Boltzmann equation was
not time-reversible; this fact, which caused much anguish, was
ultimately traced to Boltzmann's `Sto\ss zahlanzatz', or Maxwell's
`molecular chaos' hypothesis, which explicitly violates time-reversal
invariance (see \cite{LANFORD1975}).

Brown has remarked that the mathematical part of von Neumann's
measurement theory may be regarded as the first `insolubility proof'
of the quantum measurement problem. Subsequently, Fine
\cite{FINE1970}, Shimony \cite{AS1974}, Brown \cite{BROWN1986} and
Busch and Shimony \cite{PB-AS1996} have attempted, without success, to
bypass von Neumann's result by tweaking the notion of measurement; the
state vector \emph{did not collapse} in any of these attempts, whence
the claim of `the insolubility of the quantum measurement problem'.
In all cases, the apparatus was \emph{not assumed} to have a classical
description, and the quantity measured was \emph{not assumed} to be
conserved.  If collapse is a consequence of conservation laws, then
these `insolubility theorems' are only to be expected when
conservation laws are not involved.

Now assume that the quantity being measured is conserved, and
\emph{the measuring apparatus has a classical description}. Then
Sewell's measurement theory applies, and the state vector
collapses.\footnote{A textbook-level account of Sewell's theory may be
found in \cite{SEN2010}.} In Sewell's theory, the time evolution
leading to collapse is described by a Schr\"odinger equation with a
time-dependent Hamiltonian; the measurement interaction is very
short-lived. In these cases, the quantum measurement problem \emph{is
soluble}, and the need for further `interpretation' of the theory
falls away.

%%%%%%%%%%%%%%%%%%%%%%%%%%%%%%%%%%%%%%%%%%%%%%%%%%%%%%%%%%%%%%%%

\appendix

\section{Appendix: Unanswered questions}

In the last paragraph of chapter 10 of his popular book
\emph{Entanglement}, Aczel writes \cite{ACZEL2003}:

\begin{quote}
When von Neumann's seminal book appeared in English, Wigner told Abner
Shimony: ``I have learned much about quantum theory from Johnny, but
the material in his Chapter Six Johnny learnt all from me''.
\end{quote}

I wrote to Professor Shimony about this quotation. I have mislaid his
answer, but it was essentially as follows.

\begin{quote}

It is difficult to recall with certainty events that took place over
fifty years ago, but yes, Wigner did say something of the sort. I
would only question the use of the word `all'.

\end{quote}

Chapter VI of von Neumann's book is on `The Measuring Process'. The
remarks on the Compton-Simon experiment and the conclusions to be
drawn from it are in Chapter III, on `The Quantum Statistics'. Wigner
would have noticed von Neumann's misinterpretation of the
Compton-Simon experiment in no time, which gives credence to
Rechenberg's assertion, which I learned from Helmut Reeh  quite
recently. Reeh wrote:

\begin{quote}

A historian of science, H. Rechenberg, once told me that von Neumann's
book had been written over two different periods of time.  Rechenberg
intended to investigate that but died before doing so. 

\end{quote}

As is known, Wigner remained a firm believer in (i)~the strictly
quantum character of the measuring apparatus, and (ii)~the essential
role of the observer's consciousness in quantum measurement theory
\cite{EPW1963,EPW1964}. To the present author, the big question is, if
Johnny learnt some, or all, of his Chapter VI from Wigner, wasn't 
Wigner appraised of Chapter III? The error in it would not have
escaped the physicist Wigner!

Perhaps Rechenberg was right, and Wigner came into the picture only
during the writing of the second part of the book? But such questions
are best left to historians of physics to address.
 
%%%%%%%%%%%%%%%%%%%%%%%%%%%%%%%%%%%%%%%%%%%%%%%%%%%%%%%%%%%%%%%%%%%%%


\begin{thebibliography}{99}

\bibitem{ACZEL2003} Aczel, A. D. (2003). \emph{Entanglement}. The
Unlikely Story of How Scientists, Mathematicians and Philosophers
Proved Einstein's Spookiest Theory, New York, NY: Plume, A member of
the Penguin Group (USA) Inc. Authorized reprint of a hardcover edition
published by Four Walls Eight Windows, New York, NY, 2002.

\bibitem{BKS1924} Bohr, N., Kramers, H. A. and Slater, J. C. (1924).
The quantum theory of radiation, \emph{Phil.\ Mag.} Ser.\ 6, {\bf 47},
785--802. German version: \"Uber die Quantheorie der Strahlung,
\emph{Zeits.\ f.\ Phys.} {\bf 24}, 69--87 (1924).

\bibitem{BOTHE1926} Bothe, W. (1926). Absorption und Zerstreuung von
R\"ontgenstrahlen, im: \emph{Handbuch der Physik}, Vol {\bf 23}
(Quanta), 307--432. Berlin: Verlag Julius Springer.

\bibitem{BROWN1986} Brown, H. R. (1986). The insolubility proof of the
quantum-mechanical measurement problem, \emph{Found.\ Phys.} {\bf 16},
857--870.

 \bibitem{PB-AS1996} Busch, P. and Shimony, A. (1996). Insolubility
of the quantum measurement problem for unsharp observables,
\emph{Stud.\ Hist.\ Phil.\ Mod.\ Phys.} {\bf 27}, 397--404.
 
\bibitem{COMPTON1924} Compton, A. H. (1924). The scattering of
$X$-rays, \emph{J.\ Franklin Inst.} {\bf 198}, 57--72.

\bibitem{CH1924} Compton, A. H. and Hubbard, J. C. (1924).  The recoil
of electrons from scattered $X$-rays, \emph{Phys.\ Rev.} {\bf 23},
439--449.

\bibitem{CS1925} Compton, A. H. and Simon, A. W. (1925). Directed
quanta of scattered X-rays, \emph{Phys.\ Rev.} {\bf 26}, 289--299.

\bibitem{DEBYE1923} Debye, P. (1923). \emph{Phys.\ Zeits.} {\bf
} Apr.\ 15, 1923.

\bibitem{FINE1970} Fine, A. (1970). Insolubility of the quantum
measurement problem, \emph{Phys.\ Rev.} {\bf D\,2}, 2783--2787.

\bibitem{LANFORD1975} Lanford, O. E. III (1975). Time evolution of
large classical systems, in: \emph{Dynamical Systems and
Applications}, edited by J. Moser, Berlin: Springer-Verlag, Lecture
Notes in Physics, No.\ 38.

\bibitem{NM1992} Macrae, N. (1992). \emph{John von Neumann}. The
Scientific Genius Who Pioneered the Modern Computer, Game Theory,
Nuclear Deterrence, and Much More, New York: Pantheon Books. Reprinted
by the American Mathematical Society, Providence, RI, 1999.

\bibitem{SEN2010} Sen, R. N. (2010). \emph{Causality, Measurement
Theory and the Differentiable Structure of Space-Time}, Cambridge:
Cambridge University Press. Paperback edition, 2015.

\bibitem{GLS2005} Sewell, G. L. (2005). On the mathematical structure
of the quantum measurement problem, \emph{Rep.\ Math.\ Phys.} {\bf
56}, 271--290.

\bibitem{GLS2006} Sewell, G. L. (2006). Can the quantum measurement
problem be resolved within the framework of Schr\"odinger dynamics?
\emph{Markov Processes and Related Fields}, {\bf 13}, 425--440.

\bibitem{AS1974} Shimony, A. (1974). Approximate measurements in
quantum mechanics, II, \emph{Phys.\ Rev.} {\bf D\,9}, 2321--2323.
Reprinted in \cite{AS1993}.

\bibitem{AS1993} Shimony, A. (1993). \emph{Search for a
Naturalistic World View}, Vol.  II: Natural Sciences and Metaphysics,
Cambridge: Cambridge University Press.
 
\bibitem{SU1958} Ulam, S. (1958). John von Neumann, 1903--1957.
\emph{Bull.\ Amer.\ Math.\ Soc.} {\bf 64}, 1--49.

\bibitem{VN1932} von Neumann, J. (1932). \emph{Mathematische
Grundlagen der Quantenmechanik}, Berlin: Julius Springer. English
translation by Robert T. Beyer, \emph{Mathematical Foundations of
Quantum Mechanics}, Princeton: Princeton University Press, 1955
(quotations and page references refer to this edition).  New edition,
edited by Nicholas A.  Wheeler (typeset in \TeX, with Name and Subject
Indices), 2018.

\bibitem{EPW1963} Wigner, E. P. (1963). The problem of measurement,
\emph{Am.\ J. Phys.} {\bf 31}, 6--15. Reprinted in \cite{EPW1970}.

\bibitem{EPW1964} Wigner, E. P. (1964). Two kinds of reality,
\emph{The Monist} {\bf 48}, 248--264. Reprinted in ref.\
\cite{EPW1970}.

\bibitem{EPW1970} Wigner, E. P. (1970). \emph{Symmetries and
Reflections}: Scientific Essays of Eugene P. Wigner, edited by W. J.
Moore and M. Scriven, Cambridge, MA and London, England: The MIT
Press.

\bibitem{WIKIPEDIA} Wikipedia (2021). {\tt
en.wikipedia.org/wiki/John\_von\_Neumann}. Last revised 13 October
2021. 

\end{thebibliography}
\end{document}